\documentclass[12pt, draftclsnofoot,onecolumn]{IEEEtran}
\usepackage{graphicx,subcaption,xcolor,setspace,amsthm,cite,booktabs,epstopdf,dblfloatfix,blindtext,amsthm}
\usepackage[hyphens]{url}
\usepackage{algorithm,algcompatible,upgreek}
\usepackage{acro}
\usepackage[binary-units,detect-weight=true, detect-family=true]{siunitx}
\DeclareSIUnit \bitspersecond {bps}

\usepackage{amsmath,amssymb}

\newcommand{\rmm}{{\mathrm{m}}}

\makeatletter
\def\munderbar#1{\underline{\sbox\tw@{$#1$}\dp\tw@\z@\box\tw@}}
\makeatother
\newcommand{\dBm}{\SI{}{\decibel}\rmm}

\DeclareAcronym{3GPP}{
  short=3GPP,
  long=3rd generation partnership project
}
\DeclareAcronym{ADC}{
  short=ADC,
  long=analog-to-digital converter
}
\DeclareAcronym{AMP}{
  short=AMP,
  long=approximate message passing
}
\DeclareAcronym{AoA}{
  short=AoA,
  long=angle-of-arrival
}
\DeclareAcronym{AoD}{
  short=AoD,
  long=angle-of-departure
}
\DeclareAcronym{APS}{
  short=APS,
  long=azimuth power spectrum
}
\DeclareAcronym{AV}{
  short=AV,
  long=autonomous vehicle
}
\DeclareAcronym{BS}{
  short=BS,
  long=base station
}
\DeclareAcronym{BSM}{
  short=BSM,
  long=basic safety message
}
\DeclareAcronym{CP}{
  short=CP,
  long=cyclic-prefix
}
\DeclareAcronym{DFT}{
  short=DFT,
  long=discrete Fourier transform
}
\DeclareAcronym{DL}{
  short=DL,
  long=downlink
}
\DeclareAcronym{DSRC}{
  short=DSRC,
  long=dedicated short-range communication
}
\DeclareAcronym{FDD}{
  short=FDD,
  long=frequency division duplex
}
\DeclareAcronym{FMCW}{
  short=FMCW,
  long=frequency modulated continuous wave
}
\DeclareAcronym{FoV}{
  short=FoV,
  long=field-of-view
}
\DeclareAcronym{GNSS}{
  short=GNSS,
  long=global navigation satellite system
}
\DeclareAcronym{LIDAR}{
  short=LIDAR,
  long=Light detection and ranging
}
\DeclareAcronym{LOS}{
  short=LOS,
  long=line-of-sight
}
\DeclareAcronym{LPF}{
  short=LPF,
  long=low pass filter
}
\DeclareAcronym{LTE}{
  short=LTE,
  long=long term evolution
}
\DeclareAcronym{MIMO}{
  short=MIMO,
  long=multiple-input multiple-output
}
\DeclareAcronym{MRR}{
  short=MRR,
  long=medium range radar
}
\DeclareAcronym{NLOS}{
  short=NLOS,
  long=non-line-of-sight
}
\DeclareAcronym{NR}{
  short=NR,
  long=new radio
}
\DeclareAcronym{OFDM}{
  short=OFDM,
  long=orthogonal frequency-division multiplexing
}
\DeclareAcronym{ppm}{
  short=ppm,
  long=parts-per-million
}
\DeclareAcronym{RMS}{
  short=RMS,
  long=root-mean-square
}
\DeclareAcronym{RPE}{
  short=RPE,
  long=relative precoding efficiency
}
\DeclareAcronym{RSU}{
  short=RSU,
  long=roadside unit
}
\DeclareAcronym{SNR}{
  short=SNR,
  long=signal-to-noise ratio
}
\DeclareAcronym{UL}{
  short=UL,
  long=uplink
}

\DeclareAcronym{ULA}{
  short=ULA,
  long=uniform linear array
}
\DeclareAcronym{V2I}{
  short=V2I,
  long=vehicle-to-infrastructure
}
\DeclareAcronym{V2V}{
  short=V2V,
  long=vehicle-to-vehicle
}
\DeclareAcronym{V2X}{
  short=V2X,
  long=vehicle-to-everything
}
\DeclareAcronym{VRU}{
  short=VRU,
  long=vulnerable road user
}

\begin{document}
\bstctlcite{IEEEmax3beforeetal}
\title{Leveraging Sensing at the Infrastructure for mmWave Communication}
\author{Anum Ali, {\it Member, IEEE}, Nuria Gonz\'alez-Prelcic, {\it Senior Member, IEEE}, \\Robert W. Heath Jr., {\it Fellow, IEEE}, and, Amitava Ghosh, {\it Fellow, IEEE}
\thanks{This material is based upon the work supported in part by the National Science Foundation under Grant No. ECCS-1711702 and a gift from Nokia.}
\thanks{A. Ali and  R. W. Heath Jr. are with the Department of Electrical and Computer Engineering, The University of Texas at Austin, Austin, TX 78712-1687, USA \mbox{(e-mail: \{anumali,rheath\}@utexas.edu)}.}
\thanks{N. Gonz\'alez-Prelcic is with the Department of Electrical and Computer Engineering, The University of Texas at Austin, Austin, TX 78712-1687, USA, and also with the Signal Theory and Communications Department, University of Vigo, 36310 Vigo, Spain \mbox{(e-mail: ngprelcic@utexas.edu)}.}
\thanks{A. Ghosh is with Nokia Bell Labs, Naperville, IL 60563-1594 USA (e-mail: amitava.ghosh@nokia-bell-labs.com)}}
\maketitle
%
\begin{abstract}
Vehicle-to-everything (V2X) communication in the mmWave band is one way to achieve high data-rates for applications like infotainment, cooperative perception, and augmented reality assisted driving etc. MmWave communication relies on large antennas arrays, and configuring these arrays poses high training overhead. In this article, we motivate the use of infrastructure mounted sensors (which will be part of future smart cities) for mmWave communication. We provide numerical and measurement results to demonstrate that information from these infrastructure sensors reduces the mmWave array configuration overhead. Finally, we outline future research directions to help materialize the use of infrastructure sensors for mmWave communication.
\end{abstract}
\section{Introduction}\label{sec:intro}
Sensors on the next-generation vehicles will generate a large amount of data (terabytes-per-hour~\cite{Choi2016Millimeter}). The sensing ability of a single vehicle, though, is limited to \ac{LOS} and has blind spots. To overcome this limitation, the sensor data could be shared with neighboring vehicles for cooperative perception. Current vehicular communication mechanisms, i.e., \ac{DSRC} and \ac{LTE}-\ac{V2X} support use-cases that require low data-rate, e.g., cooperative maneuvers. These technologies cannot support the uses-cases that require high data-rates~\cite{3GPP22886}, e.g., infotainment, cooperative perception, and augment reality assisted driving. Millimeter-wave (mmWave) communication can overcome this limitation,  since it can support a very high data-rate owing to the high-bandwidth~\cite{Choi2016Millimeter}. 
In mmWave, a large number of antennas and directional transmission/reception is used to obtain sufficient link-margin. These large antenna arrays need to be re-configured frequently in highly dynamic channels. Link re-configuration could entail different procedures depending on the hardware architecture of the mmWave system. As an example, for phase-shifter based analog arrays, re-configuration means reselecting the best transmit and receive beam. For hybrid analog-digital architecture, re-configuration could mean determining the hybrid precoder/combiner based on updated channel information. Continous re-configuration of the antenna arrays is a source of significant training overhead. 
The usual viewpoint on the relation between sensors and communication is that communication will help with sensing by enabling the exchange of sensor data among vehicles. An alternative perspective, discussed in detail in this article, is that sensors can aid in establishing the communication link. This is possible as sensors provide information about the same physical environment that produces the wireless channel. Therefore, we outline a dual relationship between sensing and communication, where both help each other.

In this article, we start by highlighting that sensing at the infrastructure will be a part of future smart cities. Then, we discuss how infrastructure mounted sensors can help the mmWave communication system. Subsequently, we provide numerical and measurement results to demonstrate that information from infrastructure mounted sensors can reduce the mmWave array configuration overhead. Finally, we discuss some future research directions for leveraging infrastructure mounted sensors for mmWave communication.	
\section{Why sensors at the infrastructure?}\label{sec:why}
\begin{figure*}
\centering
\includegraphics[width=0.6\textwidth]{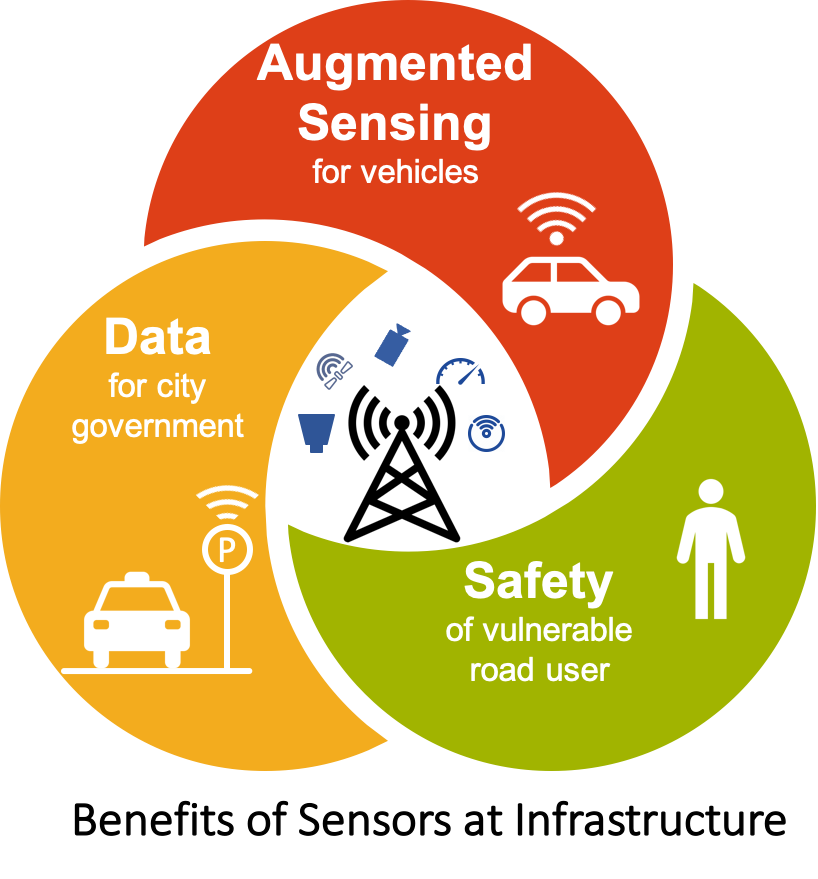}
\caption{The benefits of mounting sensors on the infrastructure include protecting the pedestrians, helping vehicles on the road by augmented sensing, and providing data for the city government.}
\label{fig:BenefitsSAI}
\end{figure*}
There are several applications of infrastructure sensing which are illustrated in Fig.~\ref{fig:BenefitsSAI}. These applications have prompted the development of smart~\acp{RSU}. NXP's \ac{RSU}~\cite{Fuller2019Smarter} is one example that performs vehicle and pedestrian detection using cameras, radars, and cloud provided information, in addition to~\ac{V2X} communication. In this section, we briefly discuss some non-communication applications of infrastructure sensors. 
\subsection{Augmented Sensing for Vehicles}
The next-generation vehicles will be equipped with several sensors, including radars, lidars, and cameras. Perception through these sensors is limited to \ac{LOS}. Cooperative perception, i.e., perception based on sensor data exchange among vehicles, can overcome the limited sensing ability of an individual vehicle. The sensors on the vehicles, however, generate a large amount of data (terabytes-per-hour~\cite{Choi2016Millimeter}), and reliable high-rate communication will be required to share this data with neighboring vehicles for cooperative perception. Therefore, the bottleneck for cooperative perception will be the market penetration of vehicles with high-rate communication capability. Initially, only a fraction of vehicles on the road will have high-rate communication capability, thus limiting the potential for cooperative perception. The smart \acp{RSU} can somewhat circumvent this limitation. The sensors on the~\ac{RSU} will have a full view of all the vehicles and other objects on the road. Thus, the~\ac{RSU} can overcome the limited sensing ability of a vehicle by sharing its own sensor data with the vehicle. This information can be used by the vehicle on the road, e.g., for better trajectory planning of an \ac{AV}~\cite{Ali2018Vehicle}.
\subsection{Safety of \acl{VRU}}
Pedestrians and bicyclists are among the most \acfp{VRU}. The \acp{VRU} are not fully protected from vehicles. The sensing ability of a vehicle is limited, and it may not detect the threat to a \ac{VRU}. In addition, if the vehicle detects the threat to a \ac{VRU}, it may lack a strategic mechanism to interact with the \ac{VRU}. This is evidenced by a recent accident involving an Uber~\acl{AV} that resulted in a  pedestrian fatality~\cite{Griggs2018How}. It is vital to have an infrastructure capable of detecting \acp{VRU} on the road. As the \acp{RSU} are typically mounted on towers or buildings, they can have what is called a \emph{bird's-eye-view} of the environment. This ensures that all the \ac{VRU} are observed/detected at the \ac{RSU}. The information about the \ac{VRU}'s position and trajectory can then be communicated to the vehicle on the road for the safety of \ac{VRU}. The \ac{RSU} may trigger a warning signal if it detects a threat to the \ac{VRU}. If the \ac{VRU} has communication capability (e.g., a smartphone), the \ac{RSU} may also inform the \ac{VRU} about the threat. 
\subsection{Data for city government}
The smart \acp{RSU} also directly provides data to the city government. The tracking of vehicles on the road can help the city in dynamically controlling the traffic flow. This not only reduces vehicle congestion but also carbon dioxide emissions. Furthermore, smart \acp{RSU} with sensing capability can find and communicate the availability of parking spots to the vehicles. This has a dual benefit of reduced time for parking spot search and better utilization of parking spots. Moreover, the data about traffic patterns and driver behaviors can be used for determining the infrastructure requirements, both in terms of new road planning and deployment of traffic signs, traffic lights and pedestrian crossings etc. A secondary impact is that the cities deploying smart infrastructure will be seen as economically competent and become attractive for new businesses. 
\section{Infrastructure sensor assisted mmWave communication}\label{sec:sensoraided}
\begin{figure}
\centering
\includegraphics[width=0.5\textwidth]{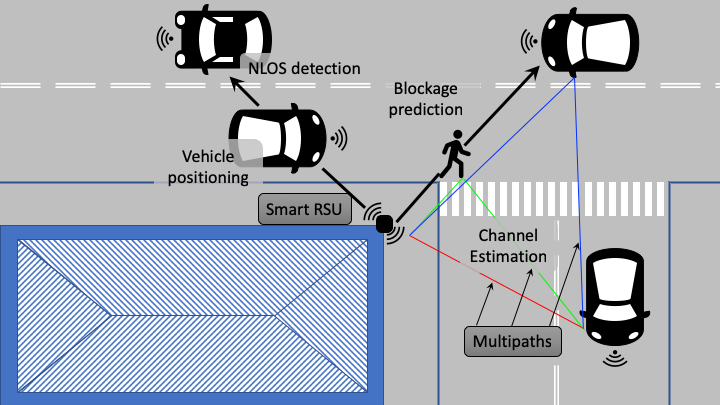}
\caption{Sensors at infrastructure can help the mmWave communication by vehicle positioning, blockage mitigation, NLOS detection, and channel estimation.}
\label{fig:SAIforComm}
\end{figure}

The sensors mounted on the infrastructure provide valuable information about the environment. The mmWave channel stems from the same environment. Therefore, it is intuitive that sensor information can be used for mmWave communication. In this section, we discuss several use cases of infrastructure mounted sensors for mmWave communication, which are illustrated in Fig.~\ref{fig:SAIforComm}.
\subsection{Vehicle positioning}
Phased arrays are suitable to keep cost and power consumption low in a large antenna system. Phased arrays are configured for directional transmission/reception by training based on codebooks that contain a certain number of beampatterns. This training entails testing all the possible pairs of transmit and receive beam patterns to determine the pair that maximizes some pre-determined criterion, e.g., highest \ac{SNR}. The beam-training process is time-consuming, especially if the codebook is large, which is typically the case for mmWave systems. The reason is that to have sufficient link margin, narrow beams with high gain have to be used. A byproduct is that a large number of beams (or codewords) is required to cover the whole field of view of the array. If position information about the vehicle is available, the number of beam patterns to be tested can be reduced~\cite{Garcia2016Location}. Typically, the position assisted strategies assume that the vehicle will obtain its position using \ac{GNSS} and this position will be conveyed to the \ac{RSU} using sub-6 GHz communication, e.g., \ac{DSRC}. There are two limitations of this approach. First, the accuracy of \ac{GNSS} based position estimates is poor, e.g., in dense urban areas. Second, the position information is shared as part of the \ac{BSM}, which is broadcasted only every \SI{100}{\milli\second}~\cite{2009J2735}. Therefore, not only the position information is inaccurate, but it may be aged as well. An alternative way is to obtain the position information directly at the \ac{RSU} using a radar~\cite{Ali2019Millimeter}. Positioning using radar is not only more accurate compared to the GNSS (depending on the parameters of the radar, e.g., the number of antennas), but also the position could be obtained and updated at a more frequent rate. 
\subsection{Blockage mitigation}
The directional nature of the mmWave communication implies vulnerability to blockage. Once the mmWave beam-training finalizes, a directional link is established, but it needs to remain unobstructed to maintain high-rate communication. It is possible for a pedestrian/vehicle to block the directional link. The power loss due to this blockage could be so severe (e.g., $\SI{30}{}-\SI{40}{\decibel}$~\cite{MacCartney2016Millimeter}) that an active link can no longer be sustained. One solution to prevent this situation is to monitor the received power, and as it starts to drop, look for alternative channel paths before the established connection is completely lost. This way, the link will be maintained by a different beam or a different \ac{RSU}. Though this approach is feasible, it is reactive in nature. Once the link quality has sufficiently dropped to detect a possible blockage, it may not take long before the link is unsustainable. An alternative approach is to use sensors mounted on the infrastructure to track the movement of pedestrians/vehicles. In this manner, potential blockages can be detected long before the received power drops substantially~\cite{Simic2016RadMAC}. This proactive method of blockage detection and beam/\ac{RSU} switching can save link failures in mmWave due to blockage.
\subsection{Channel state detection}
Sometimes it is of interest to know the state of the channel, i.e., whether the channel is \ac{LOS} or \ac{NLOS}~\cite{Huang2019Angular}. One application of this information is in channel modeling. Typically, the channel parameters are specified separately for the \ac{LOS} and \ac{NLOS} cases. As these parameters are typically derived from measurements, it is important to have the state information before the channels are classified as \ac{LOS} or \ac{NLOS}. Another application of channel state information is in positioning using communication signals, which is an active area of research. It is important to know the channel state for positioning, as positioning algorithms have a positive bias when the channel is \ac{NLOS}. 
The typical method of characterizing the channel state as \ac{LOS} or \ac{NLOS} is to use some channel characteristic, e.g., Rician K factor. There are two problems with such a strategy. One, this requires the channel knowledge, which is hard to obtain in mmWave systems. Two, the methods based on channel K-factor have a high misidentification rate~\cite{Huang2019Angular}. Alternatively, the sensors mounted on the infrastructure can track all the objects, thereby helping in distinguishing \ac{LOS} channels from \ac{NLOS} channels. As an example, lidar information and binary classification have been used to determine the channel state with high accuracy in~\cite{Klautau2019LIDAR}.
\subsection{Channel estimation}
Vehicle position, potential blockage, and channel state are channel related parameters that can be estimated using sensors. Note that as the sensors observe the same environment that produces the mmWave channel, it may be possible to directly obtain the channel information from sensors. An interesting application to this end is the estimation of mmWave channel's spatial covariance based on radar data. One approach is to put a passive radar receiver at the \ac{RSU} that captures the signals transmitted by the automotive radars of the vehicles on the road~\cite{Ali2019Passive}. Based on the radar received signals, the spatial covariance of the radar is constructed. This covariance can then be used as a proxy for mmWave channel's covariance. To use this covariance, some processing is necessary due to the differences between radar and communication channels. The radar and communication system may use a different number of antennas. In addition, radar and communication operate in different bands (e.g., \SI{73}{\giga\hertz} for communication and \SI{77}{\giga\hertz} for automotive radar). Finally, the position of the radars on the vehicles will typically be different from the communication arrays. Once the radar covariance is corrected for some (or all) of these differences it can be used in mmWave communication. As an example, in hybrid analog-digital architectures, radar covariance may be used to configure the RF-precoder. Subsequently, the baseband-precoder can be configured using in-band training. The number of RF-chains in a hybrid analog-digital system is typically much smaller than the number of antennas. Therefore, once the RF is configured, the effective system (baseband-to-baseband) has a much smaller dimension and configuring it does not pose much overhead. In phased array-based mmWave systems, the radar covariance can be used to prune the beam-pairs that are unlikely to give high \ac{SNR}.
\section{Evaluation of sensor aided mmWave communication}\label{sec:evaluation}
In this section, we provide numerical and measurement results to show the benefit of using infrastructure mounted sensors for mmWave communication. 
\subsection{Positioning at infrastructure}
\begin{figure}
\centering
\includegraphics[width=0.5\textwidth]{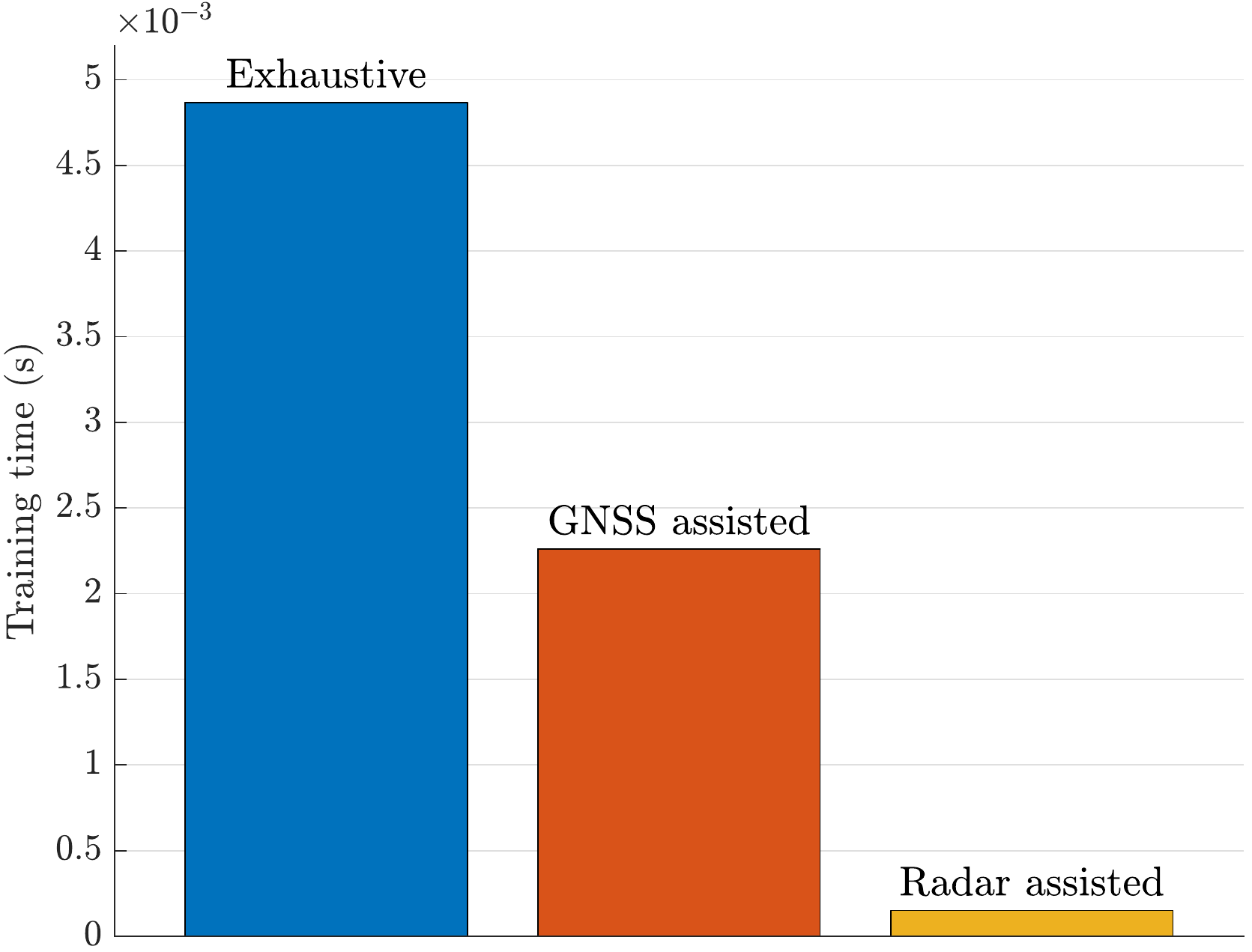}
\caption{The overhead of beam-search in phased-array based analog mmWave system with $64$ antennas at the \ac{RSU} and $16$ antennas at the vehicle. The symbol duration is $\SI{4.75}{\micro\second}$. The exhaustive beam-search requires that $1024$ beam-pairs be tested with training time $\SI{4.87}{\milli\second}$. This is reduced to $475$ beam-pairs ($\SI{2.3}{\milli\second}$) using GNSS based position information, and to $32$ beam-pairs ($\SI{0.15}{\milli\second}$) using \ac{RSU} mounted radar-based positioning. }
\label{fig:positioning}
\end{figure}

We consider the beam-search problem in a phased array-based analog mmWave system, with $64$ antennas used at the \ac{RSU} and $16$ antennas at the vehicle. Assuming DFT codebooks are used at the \ac{RSU} and the vehicle, there is a total of $64\times 16=1024$ beam-pairs that need to be tried before the best beam-pair can be determined. Assuming that each transmission symbol has a duration of $\SI{4.75}{\micro\second}$ (i.e., $\SI{4.17}{\micro\second}$ 
useful symbol duration, i.e., the shortest symbol duration in 5G-NR, and $\SI{0.58}{\micro\second}$ cyclic prefix), the total training time for exhaustive-search (i.e., testing all the beam-pairs) is $64\times16\times\SI{4.75}{\micro\second}=\SI{4.87}{\milli\second}$ as shown in Fig.~\ref{fig:positioning}. Note that, it is possible to have a hierarchical-search instead of exhaustive-search, i.e., a search over (a few) wide beams followed by a search over narrow beams (limited to the angular region of the selected wide beam). Though it might be possible to include position-information in hierarchical-search, in this paper, we focus only on exhaustive-sarch. For LOS channels, if the location of the vehicle is known, the number of beam-pairs that need to be tested can be substantially reduced. For the position information coming from \ac{GNSS}, there is an error of around $\SI{3}{\meter}-\SI{10}{\meter}$. In our simulation, we assume a $\SI{5}{\meter}$ error. Once the position information from \ac{GNSS} is available, the $\SI{5}{\meter}$ error is considered, and a subset of beam-pairs to be tested is determined. Based on the simulation, we got that the average number of beam-pairs that need to be tested with \ac{GNSS} position information is $475$. Therefore, the training time for \ac{GNSS} assisted beam-training is around $475\times\SI{4.75}{\micro\second}= \SI{2.3}{\milli\second}$. Finally, we also obtain the position information from a radar mounted on the \ac{RSU}. For radar, we calculate the expected error in radar positioning, assuming the parameters of INRAS Radarbook~\cite{INRASRadarbook}, i.e., $29$ antennas in the radar virtual array, $\SI{24}{\giga\hertz}$ operating frequency, and $10\dBm$ transmit power. Based on the results, we got that only $32$ beam-pairs need to be tested for radar, resulting in a training time of $\SI{0.15}{\milli\second}$. These results show the value of \ac{RSU} mounted radar for positioning and its subsequent use in reducing the beam-training overhead of phased-array based analog mmWave systems.
\subsection{Azimuth power spectrum estimation using radar}
For \ac{LOS} vehicles, position information can reduce the beam-training overhead considerably. For \ac{NLOS} vehicles, the position is not directly useful (though it could be in a machine learning approach). This is because it is possible that some beam not pointing in the direction of the vehicle gives higher \ac{SNR}. In such scenarios, the \ac{APS} can be used instead of the position for beam-pruning. The \ac{APS} gives the power as a function of the azimuth angles. Based on \ac{APS} it can be determined if the higher power is coming from angles other than the direction of the vehicle. The \ac{APS} can be recovered from the covariance of a passive radar at the \ac{RSU}.
 
\begin{figure}
\centering
\includegraphics[width=0.5\textwidth]{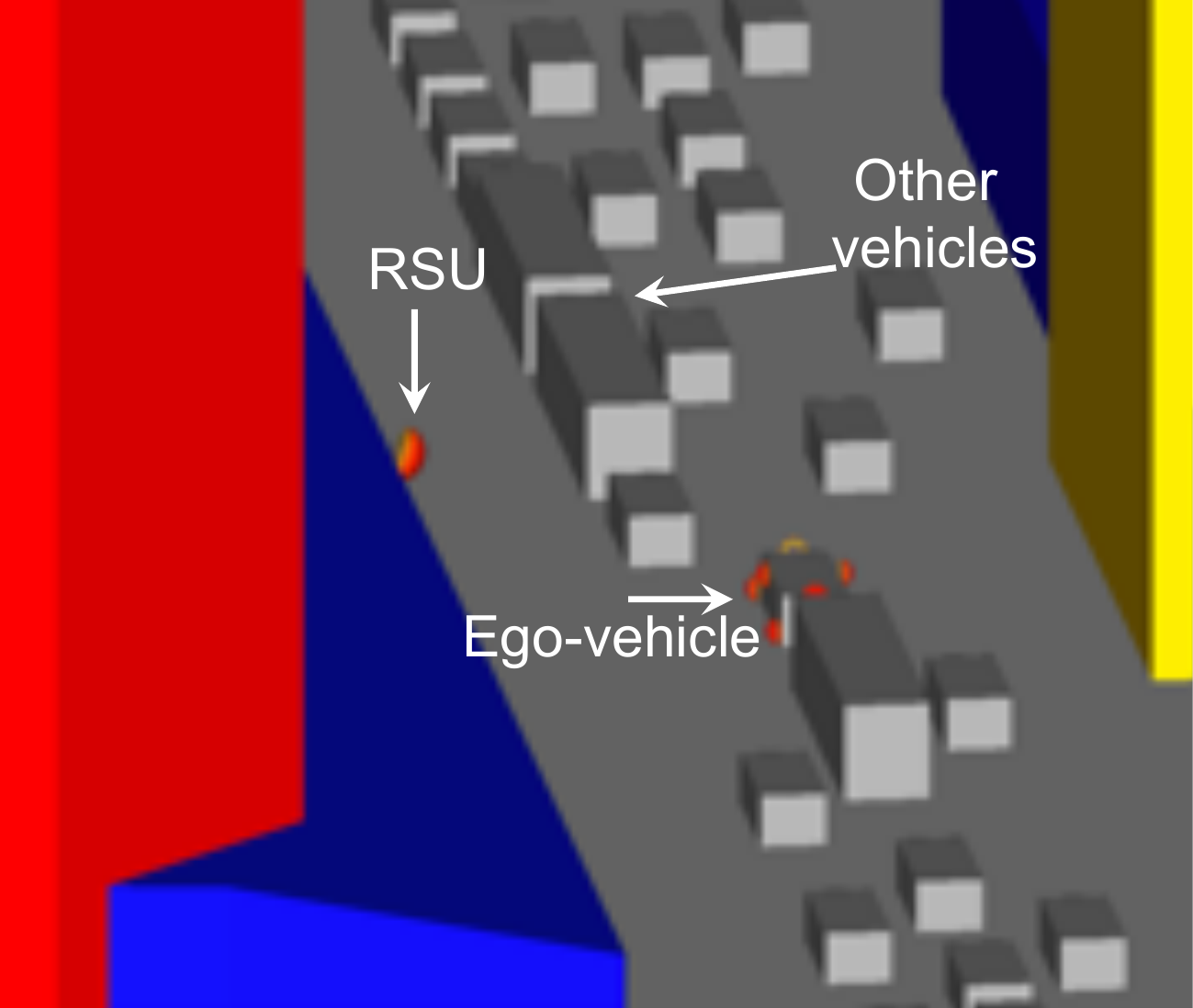}
\caption{The ray-tracing scenario with randomly dropped vehicles on the road. The buildings are made of concrete, the road is made of asphalt, and the randomly dropped vehicles on the road are made of metal. The inter-vehicle spacing, the position of communication arrays, the height of RSU, and the road width are consistent with 3GPP-V2X evaluation methodology~\cite{3GPP37885}.}
\label{fig:insite}
\end{figure}

\begin{figure}[h!]
\centering
        \includegraphics[width=0.5\textwidth]{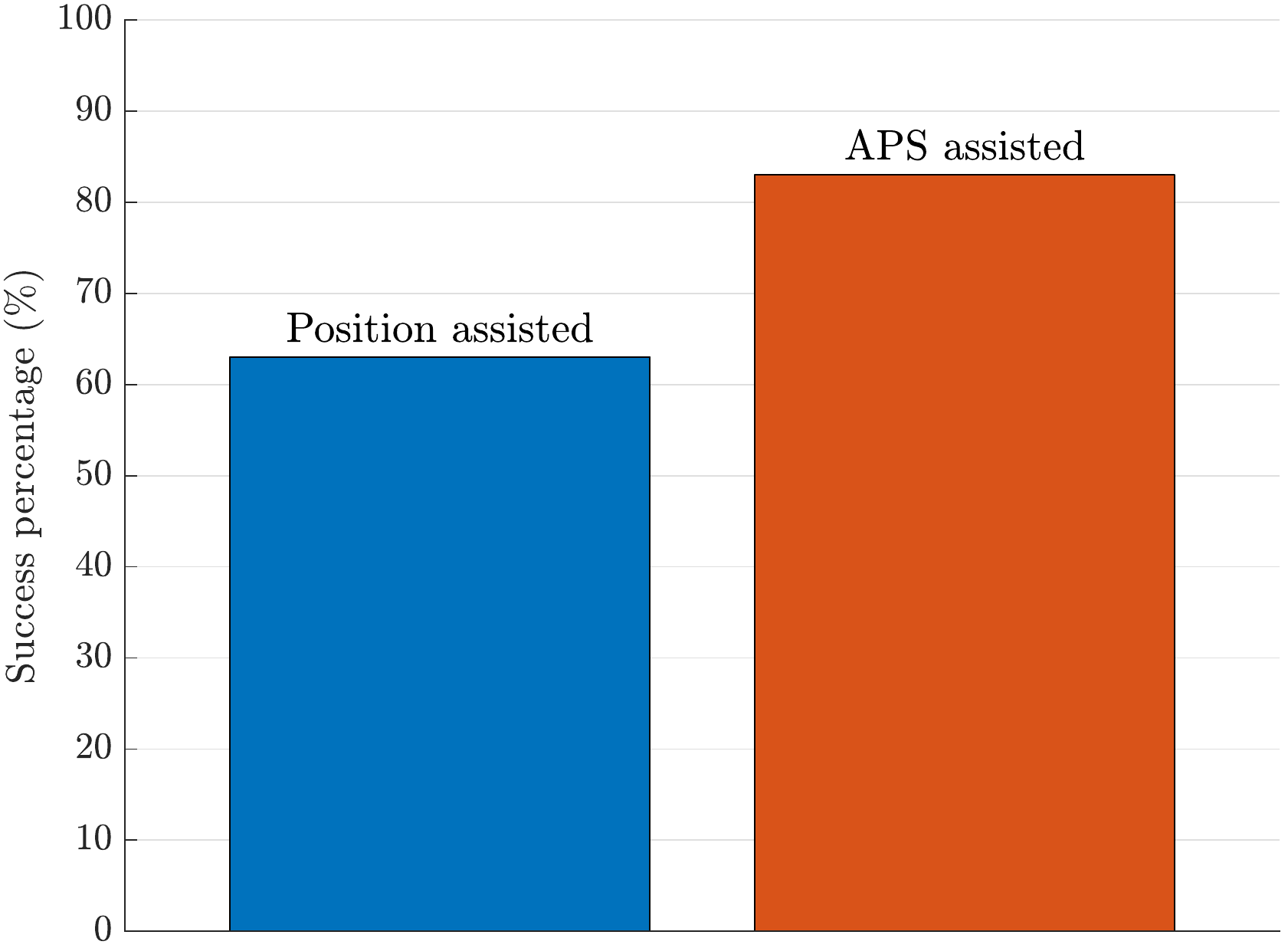}
\caption{Success percentage (i.e., the percentage of time the same beam-pair is recovered as exhaustive-search) of \ac{APS} assisted strategy and position-assisted strategy. The $83\%$ success for \ac{APS}-assisted in comparison with $63\%$ success of position-assisted shows that \ac{APS} information recovered from passive radar at the \ac{RSU} is more useful in \ac{NLOS} scenarios than the position information.}
        \label{fig:PercResults}
\end{figure}

To evaluate the \ac{APS} assisted strategy, we use Wireless Insite ray-tracing simulations. We use the setup shown in Fig.~\ref{fig:insite}. The buildings are made of concrete, the road is made of asphalt, and the randomly dropped vehicles on the road are made of metal. Some parameters, e.g., the inter-vehicle spacing, the position of communication arrays, the height of RSU, and the road width are consistent with 3GPP-V2X evaluation methodology~\cite{3GPP37885}. The radar operates at $\SI{76}{\giga\hertz}$ and the communication system operates at $\SI{73}{\giga\hertz}$. Both the radar and the communication transmitter on the \ac{RSU} have $128$ antennas with $4$ arrays of $16$ antennas each on the vehicle. We use success percentage as a metric to compare the proposed \ac{APS} assisted strategy with position-assisted strategy. The success percentage is the percent of times the proposed strategy recovers the same beam-pair as exhaustive-search. The results of this experiment are shown in Fig.~\ref{fig:PercResults}. To get the best beam-pair exhaustive-search requires $128\times 16 = 2048$ training symbols. Whereas in our experiments, the position-assisted strategy required only $560$ training symbols. For a fair comparison, we also used the same number of training symbols for \ac{APS} assisted strategy. The position-assisted strategy, however, recovers the best beam-pair only $63\%$ of the time, whereas \ac{APS} assisted strategy recovers the best beam-pair $83\%$ of the time as shown in Fig.~\ref{fig:PercResults}. This result shows the value of passive radar at the \ac{RSU} to reduce the beam-training overhead in \ac{NLOS} scenarios.
\subsection{Similarity of radar and communication \ac{APS} in measurements}
\begin{figure*}[h!]
\centering
\begin{subfigure}{0.45\textwidth}
        \centering
        \includegraphics[width=\textwidth]{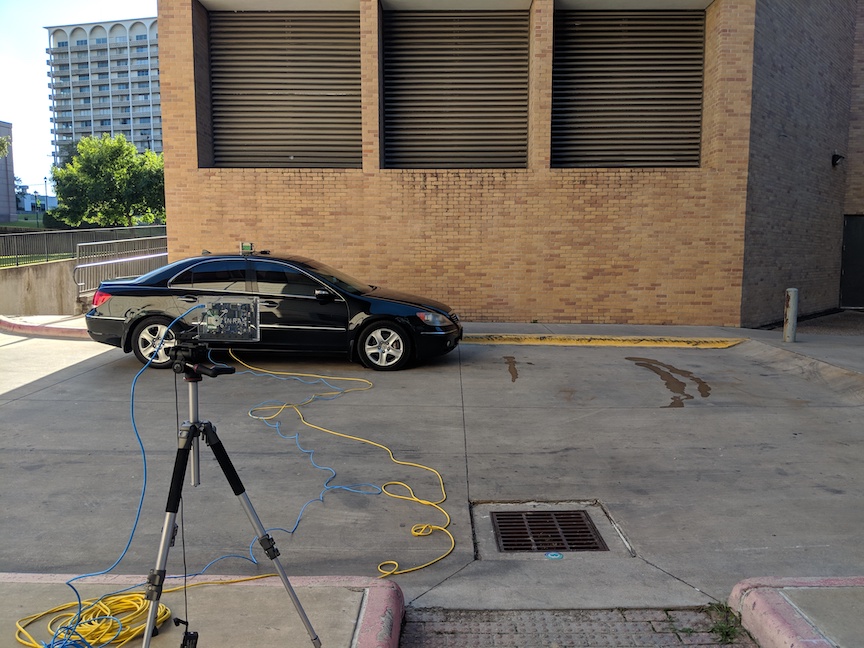}
        \caption{First measurement location.}
        \label{fig:ScenarioA}
        \end{subfigure}
            \begin{subfigure}{0.45\textwidth}
        \centering
        \includegraphics[width=\textwidth]{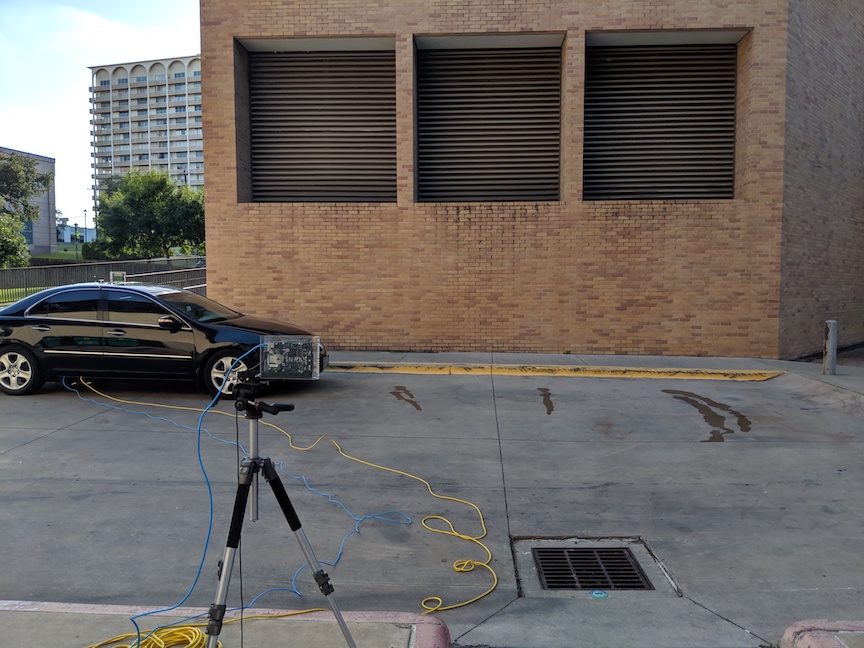}
        \caption{Second measurement location.}
        \label{fig:ScenarioB}
    \end{subfigure}
  \hfill\\
  \begin{subfigure}{0.45\textwidth}
        \centering
        \includegraphics[width=0.6\textwidth]{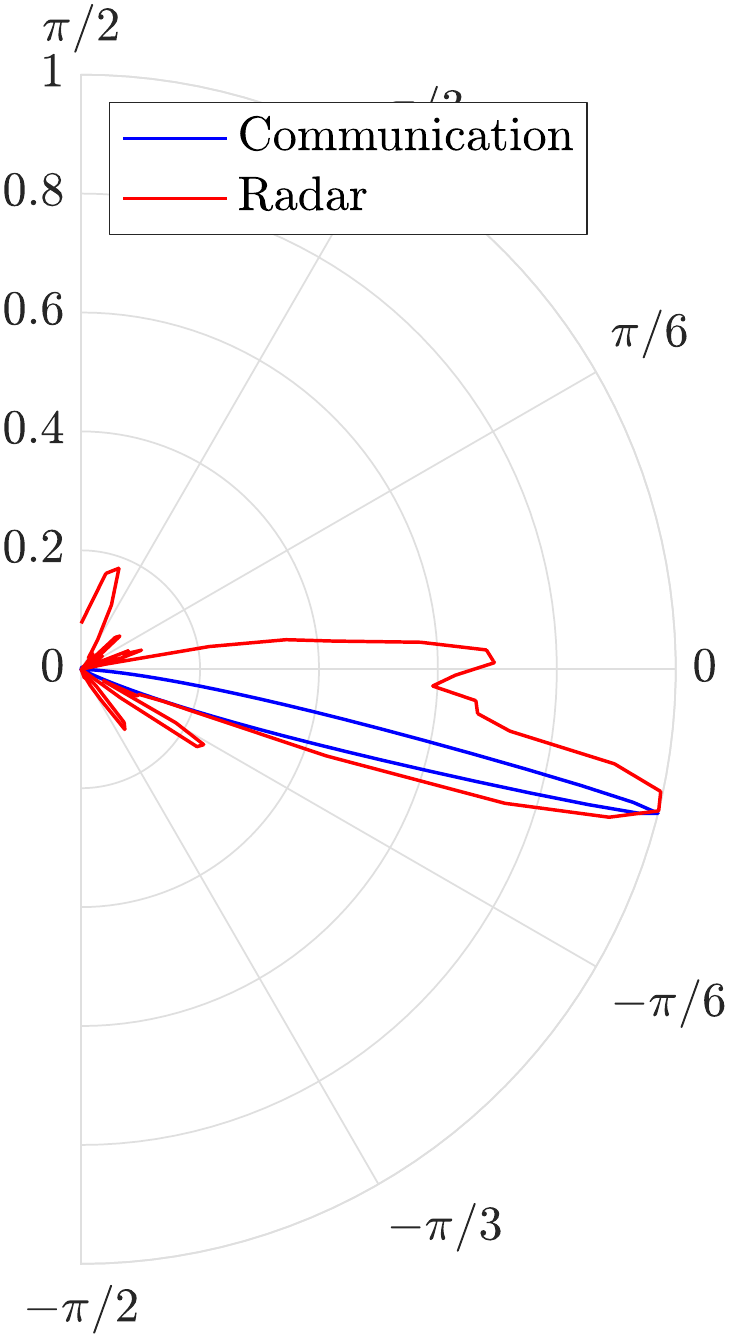}
        \caption{\ac{APS} for first measurement location.}
        \label{fig:ScenarioResA}
        \end{subfigure}
            \begin{subfigure}{0.45\textwidth}
        \centering
        \includegraphics[width=0.6\textwidth]{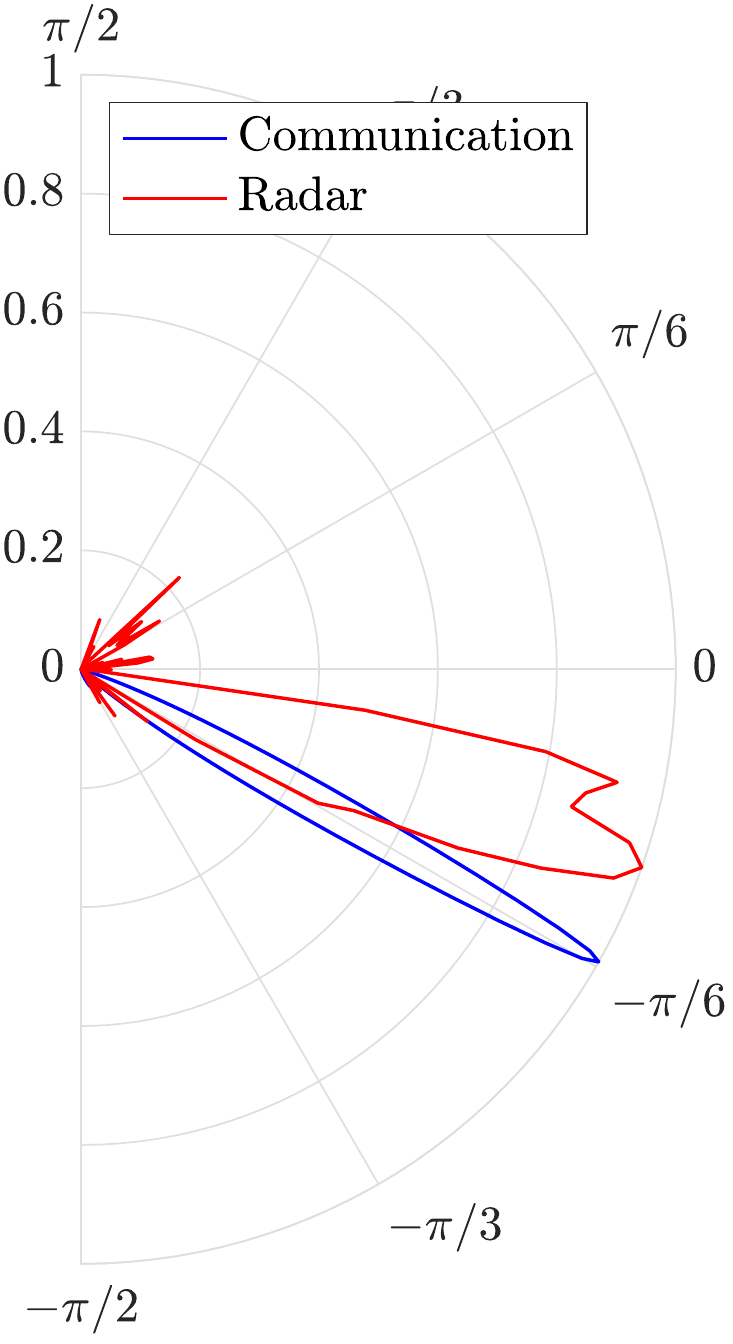}
        \caption{\ac{APS} for second measurement location.}
        \label{fig:ScenarioResB}
    \end{subfigure}
    \hfill
        \caption{The measurement scenario with radar and communication transmitter mounted on a tripod. The communication receiver is mounted on the vehicle. The normalized \ac{APS} of the radar and the communication shows significant similarity in radar and communication channel.}
        \label{fig:ScenarioandRes}
\end{figure*}

The results presented in the previous subsection used passive radar at the \ac{RSU} (that taps the signals transmitted from the automotive radars on the vehicle) to reduce the beam-training overhead in \ac{NLOS} scenarios. We now present the measurement results to show that the \ac{APS} of active mono-static radar is similar to the communication channel's \ac{APS} in \ac{LOS} scenarios. The measurement setup is shown in Fig.~\ref{fig:ScenarioandRes}. An active INRAS Radarbook~\cite{INRASRadarbook} is mounted on the tripod as shown in Fig.~\ref{fig:ScenarioA} and Fig.~\ref{fig:ScenarioB}. We modified another Radarbook to turn off its transmission function. This modified Radarbook was placed on the vehicle and served as the receiver for the communication system. Two example instances of this configuration are shown in Fig.~\ref{fig:ScenarioA} and Fig.~\ref{fig:ScenarioB}. The normalized \acp{APS} of the radar and communication channel are shown in Fig.~\ref{fig:ScenarioResA} and Fig.~\ref{fig:ScenarioResB}. The \acp{APS} are normalized to be unity in the maximum gain direction. We can see from Fig.~\ref{fig:ScenarioResA} that the dominant direction of the radar and communication \ac{APS} is completely aligned. This implies that the angular information extracted from an active mono-static radar mounted on the \ac{RSU} can be used to configure the mmWave link. There is a slight mismatch in the dominant direction of radar and communication in Fig.~\ref{fig:ScenarioResB}. This is because the communication transmitter is mounted at the center of the vehicle roof. In comparison, radar \ac{APS} depends on the dominant reflecting point on the vehicle, which in this case is the front of the car. This results in a slight mismatch in the dominant direction of radar and communication \ac{APS}. Nevertheless, from these measurements, it can be discerned that the angular information provided by radar is highly correlated with the communication channel and can be used to reduce the training overhead considerably.

\section{Future research directions}
There are several directions for future work in leveraging infrastructure sensors for communication. We highlight some of these directions below.
\subsection{Redesigning sensors}
The radar sensors on vehicles are usually mounted just above the bumper. When the same sensors are deployed on the \ac{BS}, performance suffers due to the restricted \acl{FoV}, different scattering assumptions, higher distance from the ground, and the potential for mutual interference among multiple~\ac{LOS}~\acp{BS}. As a result, radar and other sensors may need to be redesigned with a cellular infrastructure deployment in mind. 
\subsection{Data fusion}
The data from multiple sensors may be fused to improve situational awareness for automated driving applications. The sensors though may also be fused to help with mmWave communication, not environmental perception. It is interesting to develop techniques for mmWave communication that can cope with different sensing modalities, including in particular the diverse rates at which the sensors produce data. 
\subsection{Mathematical frameworks for exploiting sensor information}
There are several opportunities to use the sensor information for mmWave communication. As discussed earlier, preemptive \ac{RSU} or beam switching for blockage mitigation, \ac{LOS} detection, positioning for beam-training overhead reduction, and covariance or \ac{APS} estimation are but a few example cases where the sensors can help the mmWave communication. For each of these applications, several strategies can be employed to use sensor information for mmWave communication. Consider the example of covariance estimation, which can be performed using a compressed sensing framework. In compressed sensing, side information can be used through weighted compressed sensing, where the sparse prior is non-uniform. Similarly, channel estimation can be performed using an approximate message passing framework (i.e., a class of algorithms for linear inverse problems and their generalizations), that permits the use of side-information. Finally, machine learning strategies can be used to uncover correlations between the mmWave communication channel and the radar signatures.
\subsection{Multi-user scenario}
The results presented in this article were link-level, i.e., for a single link between a vehicle and a \ac{RSU}. It is necessary to extend the proposed strategies to a multi-user scenario. As an example, for positioning applications, the accuracy of the radar positioning may decrease in the presence of multiple targets. Similarly, for passive radar-based \ac{APS} estimation, the transmissions from automotive radars of multiple vehicles will overlap. Therefore, intelligent strategies will be needed to use the radar information for mmWave links in such multi-user scenarios.
\section{Conclusions}\label{sec:conc}
Infrastructure mounted sensors have the potential to dramatically enhance mmWave communications, in important applications of communications in automated and cyber-physical systems.  When co-located with edge-computing,  such sensors can be fused to provide a bird's-eye-view that is beneficial for automated driving. Fully exploiting infrastructure mounted sensors for mmWave requires developing sensing hardware for \ac{BS} deployments and new algorithms that can learn from this data to recommend appropriate communication actions. While we emphasized ground vehicles in this article, it should be clear that the proposed sensing solution is promising as well for aerial vehicles and robotics. In those applications, where batteries may be small and compute capabilities limited, the remote sensing and computation change design constraints. Now is the time for sensing and communication to come together. 
\bibliographystyle{IEEEtran}
\bibliography{Abbr,Sensing_at_the_Infrastructure}{}
\end{document}